\documentclass[11pt]{article} 
\pdfoutput=1 
\begin{document} 
\title{Vorticity and Capillaries at the Surface of a Jet} 
\author{Matthieu A. Andre and Philippe M. Bardet \\ 
\\\vspace{6pt} Mechanical and Aerospace Department \\ The George Washington University, Washington, DC, USA} 
\maketitle 

Separated liquid-gas flows are encountered in countless environmental and industrial contexts. The interface between a liquid and a gas deforms to accommodate the flow on both sides and can lead to dramatic behaviors such as sprays, air entrainment, phase change, etc. Free surface instabilities arise when the surface dynamics is solely controlled by the liquid phase and small initial disturbances in the flow result in surface deformation.  One particular mechanism appears when a wall-bounded flow with thin laminar boundary layer transitions to a free-surface flow, such as at exit of liquid jets or on a cavitating body. The instability stems from an inflexion point in the boundary layer mean velocity profile; this profile results from the transition from a no-slip to a shear-free boundary condition [1]. 

This study aims at understanding mechanisms governing this instability and at characterizing the flow evolution. The experiment consists of a 2D water jet ($20.3mm$ deep by $146mm$ wide) flowing into a rectangular open channel at velocity ranging from $1$ to $10m/s$ ($Re$= $6.8 \times 10^3$ to $6.8 \times 10^4$) [2]. This canonical geometry is studied with non-intrusive diagnostics allowing good characterization of the phenomena.  Only the first $3mm$ below the surface are investigated.

Experimental data showed in the fluid dynamics video consists of: 1- surface observations made with flash photography and Planar Laser Induced Fluorescence (PLIF) and 2- simultaneous surface profile and velocity field obtained with PLIF and Particle Image Velocimetry (PIV). The small temporal and spatial scales associated with the flow (sub-milliseconds and millimeters) necessitate high-speed and high-magnification imaging. The highly deformed interface cannot be reliably masked on the raw PIV images alone; PLIF imaged from above the layer is used for profilometry and to generate interface masks for PIV processing. Both PIV and PLIF images are calibrated simultaneously and corrected for perspective distortions in order to be used jointly.

The formation of capillary waves resulting from the shear layer roll up was first suggested by Brennen [1]; here this is confirmed with PLIF.  Additionally, downstream of steep capillaries, injection of counter rotating vortex pairs in the bulk of the flow has been identified. Time-resolved PIV measurements reveal the roll up of the shear layer (negative vorticity), and the presence of positive surface vorticity in the sharp troughs of the capillaries as derived by Longuet-Higgins [3] for potential flows, though. Because this secondary surface vorticity is confined to a thin surface layer, it is not always resolved with PIV. A stable case appears when the vortices stay in phase with the waves sustaining them for tens of wavelengths. However, as the waves grow in amplitude, they become non-linear and collide. Collisions inject the positive vorticity from the surface layer into the bulk of the flow. The positive vorticity then interacts with the negative vorticity from the shear layer leading to the formation of an asymmetrical counter rotating vortex pair. This pair translates and rotates according to their mutually induced velocities.  As a consequence of the vorticity transfer, the free surface curvature strongly decreases. In this process air can also be entrained. The interaction between the vortex pair and the entrained bubble still has to be resolved experimentally.

This study shows the role that this type of instability can have on interphase transfers especially through surface renewal as well as on phases mixing via air entrainment.  Farther downstream of these events, water ligaments and droplets have been observed, but are not reported here.

\section*{References} 
[1] Brennen, C. E., 1970, “Cavity Surface Wave Patterns and General Appearance,” J. Fluid Mech. 44 ptI, pp. 33-49 

\noindent
[2] Andre, M.A. and Bardet, P.M., 2012, “Experimental investigation of boundary layer instabilities on the free surface of non-turbulent jet”, Proceedings of the Open Forum on Multiphase Flows, FEDSM2012-72328

\noindent
[3] Longuet-Higgins, M.S., 1992, “Capillary rollers and bores”, J. Fluid Mech. 240, pp. 659-679

\end{document}